\documentclass{PoS}

\title{Non-linear QCD dynamics and exclusive production in $ep$ collisions}

\ShortTitle{Non-linear QCD dynamics and exclusive production}

\author{\speaker{M.V.T. Machado}\\
         High Energy Physics Phenomenology Group, GFPAE  IF-UFRGS \\
Caixa Postal 15051, CEP 91501-970, Porto Alegre, RS, Brazil\\
        E-mail: \email{magnus@if.ufrgs.br}}

\author{V.P. Gon\c{c}alves and A.R. Meneses\\
 Instituto de F\'{\i}sica e Matem\'atica, Universidade Federal de
Pelotas\\  Caixa Postal 354, CEP 96010-900, Pelotas, RS, Brazil.}

\abstract{ In this contribution we analyse the cross sections for the exclusive vector meson production as well as the deeply virtual Compton scattering (DVCS) relying on the color dipole approach and considering the numerical solution of the  Balitsky-Kovchegov equation including running coupling corrections. Comparisons to DESY-HERA data on exclusive processes and predictions to the Large Hadron Electron Collider, LHeC, are presented.}

\FullConference{35th International Conference of High Energy Physics - ICHEP2010,\\
		July 22-28, 2010\\
		Paris France}

\begin{document}
\section{Introduction}
At high energies, the growth of the parton distribution is expected to saturate, forming a  Color Glass Condensate (CGC), whose evolution with energy is described by an infinite hierarchy of coupled equations for the correlators of  Wilson lines \cite{CGC}.
In the mean field approximation, the first equation of this  hierarchy decouples and boils down to a single non-linear integro-differential  equation: the Balitsky-Kovchegov (BK) equation \cite{BK}. This equation determines, in the large-$N_c$ (the number of colors) limit, the evolution of the two-point correlation function, which corresponds to the  scattering amplitude ${\cal{N}}(x,r,b)$ of a dipole off the CGC, where $r$ is the dipole size and $b$ the impact parameter. This quantity  encodes the information about the hadronic scattering  and then about the non-linear and quantum effects in the hadron wave function. Recently, the next-to-leading order corrections to BK equation were calculated  \cite{BKNLO} through the ressumation of $\alpha_s N_f$ contributions to all orders, where $N_f$ is the number of flavors. Such calculation allows one to estimate the soft gluon emission and running coupling corrections to the evolution kernel and it was found that the dominant contributions come from the running coupling corrections, which allows to determine the scale of the running coupling in the kernel. More recently, a global analysis of the small $x$ data for the proton structure function using the improved BK equation was performed \cite{bkrunning}. In contrast to the  BK  equation at leading logarithmic $\alpha_s \ln (1/x)$ approximation, which  fails to describe data, the inclusion of running coupling effects to evolution renders BK equation compatible with them.

Exclusive processes in deep inelastic scattering (DIS) have appeared as key reactions to trigger the generic mechanism of diffractive scattering.
In particular, the diffractive vector meson production and deeply virtual Compton scattering (DVCS) have been extensively studied at HERA and provide a valuable  probe of the  QCD dynamics at high energies. In a general way, these processes are driven by the gluon content of target (proton or nuclei) which is strongly subject to parton saturation effects as well as considerable nuclear shadowing corrections when one considers scattering on nuclei. In particular, the cross section for exclusive processes in DIS are proportional to the square of scattering amplitude, which turn it  strongly sensitive to the underlying QCD dynamics.  In this contribution we summarize the main results published in Ref. \cite{GMM} where we make use of numerical solution of the  BK equation including running coupling corrections (BK RC) in order to estimate the contribution of the saturation physics for exclusive processes. Our analysis is  relevant for the physics to be studied  in future electron - proton collider, as e.g. the LHeC \cite{dainton}. In next section we provide the main formula for computing the differential cross section for exclusive processes in DIS and  discuss the numerical results considering the solution of the BK RC model.

\section{Exclusive processes in DIS and the RC BK solution}
\label{exc}

Let us consider photon-hadron scattering in the dipole frame, in which most of the energy is
carried by the hadron, while the  photon  has
just enough energy to dissociate into a quark-antiquark pair
before the scattering. In this representation the probing
projectile fluctuates into a
quark-antiquark pair (a dipole) with transverse separation
$r$ long after the interaction, which then
scatters off the hadron.
In the dipole picture the   amplitude for production of an exclusive final state $E$, such as a vector meson ($E = V$) or a real photon in DVCS ($E = \gamma$) is given by:
\begin{eqnarray}
 {\cal A}_{T,L}^{\gamma^*p \rightarrow E p}(x,Q^2,\Delta)  =  i
\int dz \, d^2r \, d^2b  e^{-i[b-(1-z)r].\Delta}  (\Psi_{E}^* \Psi)_T \,2 {\cal{N}}(x,r,b)
\label{sigmatot2}
\end{eqnarray}
where  the factor $[i(1-z)r].\Delta$ in the exponential  arises when one takes into account non-forward corrections to the wave functions. Therefore, the differential cross section  for  exclusive production is given by
\begin{eqnarray}
\frac{d\sigma_{T,L}}{dt} (\gamma^* p \rightarrow E p) = \frac{1}{16\pi} |{\cal{A}}_{T,L}^{\gamma^*p \rightarrow E p}(x,Q^2,\Delta)|^2\,(1 + \beta^2)\,,
\label{totalcs}
\end{eqnarray}
where $\beta$ is the ratio of real to imaginary parts of the scattering
amplitude. For the case of heavy mesons, skewness corrections are quite important and they are also taken  into account. For the meson wavefunction, we have considered \cite{GMM} the Gauss-LC  model which is a simplification of the DGKP wavefunctions.  In photoproduction, this leads only to an  uncertainty  of a few percents in overall normalization. We consider the quark masses $m_{u,d,s} = 0.14$ GeV, $m_c = 1.4$ GeV and $m_b=4.5$ GeV.

The scattering amplitude ${\cal{N}}(x,r,b)$   contains all
information about the target and the strong interaction physics.
In the Color Glass Condensate (CGC)  formalism \cite{CGC}, it  encodes all the
information about the
non-linear and quantum effects in the hadron wave function.  In leading
order (LO), and in the translational invariance approximation---in which the scattering
amplitude does not depend on the collision impact parameter $b$---it reads

	\begin{eqnarray}\label{eq:bklo}
		\frac{\partial {\cal{N}}(r,Y)}{\partial Y} & = & \int {\rm d}r_1\, K^{\rm{LO}}
		(r,r_1,r_2)
		[{\cal{N}}(r_1,Y)+{\cal{N}}(r_2,Y) -  {\cal{N}}(r,Y)-{\cal{N}}(r_1,Y){\cal{N}}(r_2,Y)], \nonumber \\
\label{eq:klo}
		K^{\rm{LO}}(r,r_1,r_2) & = &  \frac{N_c\alpha_s}{2\pi^2}\frac{r^2}{r_1^2r_2^2},
	\end{eqnarray}
where ${\cal{N}}(r,Y)$ is the scattering amplitude for a dipole (a quark-antiquark pair)
off a target, with transverse size $r$, $Y\equiv \ln(x_0/x)$ ($x_0$ is the value of $x$ where the evolution starts), and $r_2 = r-r_1$. This equation is a
generalization of the linear BFKL equation (which corresponds of the first three terms), with the inclusion
of the (non-linear) quadratic term, which damps the indefinite growth of the amplitude
with energy predicted by BFKL evolution.
\begin{figure}[t]
\centering
\includegraphics[scale=0.35]{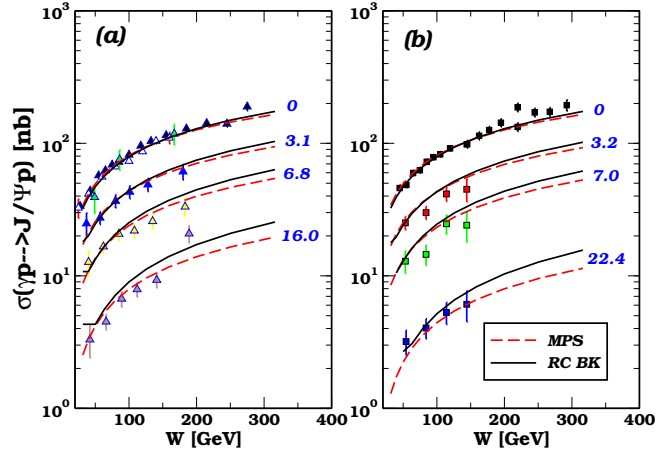}
\caption{Energy dependence of the $\gamma p$ cross section for $J/\Psi$ production  for different photon virtualities. Data from (a) ZEUS and (b) H1 collaborations (see Ref. \cite{GMM} and references therein).}
\label{fig:1}
\end{figure}
From recent numerical studies of the improved BK equation, it has been confirmed that the running coupling corrections lead to a considerable
increase in the anomalous dimension and to a slow-down of the evolution
speed, which implies, for example, a slower growth of the saturation scale with
energy, in contrast with the faster growth predicted by the LO BK equation. Moreover,
as shown in \cite{bkrunning} the improved BK equation has been shown to
be really successful when applied to the description of the $ep$ HERA data for the inclusive proton structure function.

In what follows we show some results of RC BK approach compared to the experimental data on exclusive processes at DESY-HERA and present our predictions. As a sample of studies presented in Ref. \cite{GMM}, Fig. \ref{fig:1} presents the  predictions of the RC BK model for the diffractive $J/\Psi$ production and compare them to the ZEUS (a) and H1  (b) data \cite{GMM}.   The data description is reasonable since it is a parameter-free calculation. The contribution of real part of amplitude increase by 10 \%  the overall normalization, while the skewedness have a 20 \% effect \cite{GMM}. For comparison, we show also the results for the Marquet-Peschanski-Soyez model \cite{MPS} (MPS) for the dipole cross section.

Let us now perform a preliminary study for the LHeC machine \cite{dainton}. Using the design with a electron beam having laboratory energy of $E_e=140$ GeV, the center of mass energy will reach $W_{\gamma p}=1.4$ TeV and a nominal luminosity of order $10^{33}$ cm$^{-2}$s$^{-1}$. Our estimates for the exclusive vector meson photoproduction cross sections are the following. One gets $\sigma(\gamma +p\rightarrow V+p)\simeq 18$ $\mu$b, 503 nb and 3.4 nb for $\rho^0$, $\psi(1S)$ and $\Upsilon (1S)$, respectively. The average error is of order 10\%, which includes uncertainties on the meson wavefunctions, elastic slope and skewedness effects. For DVCS, we estimate the value $\sigma (\gamma^*+p\rightarrow \gamma+p)\simeq 28$ nb at photon virtuality of $Q^2=8$ GeV$^2$, with a few percent error due to elastic slope  and skwedness modeling.

As a summary, we presented an analysis of exclusive production in small-$x$ deep inelastic scattering in terms of the non-linear QCD dynamics. This approach was performed using the recent calculation of the running coupling corrections to the BK equation.  We find a fairly good agreement with experimental data using a parameter-free calculation and provide predictions to the LHeC experiment. Our predictions are relevant for the physics programs in the ongoing experiment LHeC and  in the photoproduction processes in coherent proton - proton interactions at the LHC.

\end{document}